\documentclass[%
 reprint,
superscriptaddress,
 amsmath,amssymb,
 aps,
]{revtex4-2}

\usepackage{graphicx}
\usepackage{dcolumn}
\usepackage{bm}
\usepackage{hyperref}
\usepackage{gensymb} 

\usepackage[dvipsnames]{xcolor}   

\begin{document}

\preprint{APS/123-QED}

\title{Radiative Decay of the $^{229m}$Th Nuclear Clock Isomer in Different Host Materials}

\author{S.\,V. Pineda}

\email{skyy.pineda@kuleuven.be}
\affiliation{KU Leuven, Instituut voor Kern- en Stralingsfysica, Celestijnenlaan 200D, 3001 Leuven, Belgium}
\author{P. Chhetri}%
\email{pchhetri@uni-mainz.de}
\affiliation{KU Leuven, Instituut voor Kern- en Stralingsfysica, Celestijnenlaan 200D, 3001 Leuven, Belgium}
\affiliation{Department Chemie - Standort TRIGA, Johannes Gutenberg-Universität Mainz, D-55099 Mainz, Germany}
\author{S. Bara}
\affiliation{KU Leuven, Instituut voor Kern- en Stralingsfysica, Celestijnenlaan 200D, 3001 Leuven, Belgium}
\author{Y. Elskens}
\affiliation{KU Leuven, Instituut voor Kern- en Stralingsfysica, Celestijnenlaan 200D, 3001 Leuven, Belgium}
\author{S. Casci}
\affiliation{KU Leuven, Instituut voor Kern- en Stralingsfysica, Celestijnenlaan 200D, 3001 Leuven, Belgium}
\author{A.N. Alexandrova}
\affiliation{Department of Chemistry and Biochemistry, University of California, Los Angeles, CA 90095, USA}
\author{M. Au}
\affiliation{CERN SY-STI, Espl. des Particules 1, Geneva, 1211, Switzerland}
\author{M. Athanasakis-Kaklamanakis}
\affiliation{KU Leuven, Instituut voor Kern- en Stralingsfysica, Celestijnenlaan 200D, 3001 Leuven, Belgium}
\affiliation{CERN, Experimental Physics Department, CH-1211 Geneva 23, Switzerland}
\affiliation{Blackett Laboratory, Centre for Cold Matter, Imperial College London, SW7 2AZ London, United Kingdom}
\author{M. Bartokos}
\affiliation{Vienna Center for Quantum Science and Technology, Atominstitut, TU Wien, 1020 Vienna, Austria}
\author{K. Beeks}
\affiliation{LUMES, Institute of Physics, EPFL, Lausanne CH-1015, Switzerland}
\author{C. Bernerd}
\affiliation{KU Leuven, Instituut voor Kern- en Stralingsfysica, Celestijnenlaan 200D, 3001 Leuven, Belgium}
\affiliation{CERN SY-STI, Espl. des Particules 1, Geneva, 1211, Switzerland}
\author{A. Claessens}
\affiliation{KU Leuven, Instituut voor Kern- en Stralingsfysica, Celestijnenlaan 200D, 3001 Leuven, Belgium}
\author{K. Chrysalidis}
\affiliation{CERN SY-STI, Espl. des Particules 1, Geneva, 1211, Switzerland}
\author{T.E. Cocolios}
\affiliation{KU Leuven, Instituut voor Kern- en Stralingsfysica, Celestijnenlaan 200D, 3001 Leuven, Belgium}
\author{J.G. Correia}
\affiliation{Centro de Ciências e Tecnologias Nucleares, Instituto Superior Técnico, Universidade de Lisboa, 2695-066 Bobadela LRS, Portugal}
\author{H. De Witte}
\affiliation{KU Leuven, Instituut voor Kern- en Stralingsfysica, Celestijnenlaan 200D, 3001 Leuven, Belgium}
\author{R. Elwell}
\affiliation{Department of Physics and Astronomy, University of California, Los Angeles, CA 90095, USA}
\author{R. Ferrer}
\affiliation{KU Leuven, Instituut voor Kern- en Stralingsfysica, Celestijnenlaan 200D, 3001 Leuven, Belgium}
\author{R. Heinke}
\affiliation{CERN SY-STI, Espl. des Particules 1, Geneva, 1211, Switzerland}
\author{E.R. Hudson}
\affiliation{Department of Physics and Astronomy, University of California, Los Angeles, CA 90095, USA}
\author{F. Ivandikov}
\affiliation{KU Leuven, Instituut voor Kern- en Stralingsfysica, Celestijnenlaan 200D, 3001 Leuven, Belgium}
\author{Yu. Kudryavtsev}
\affiliation{KU Leuven, Instituut voor Kern- en Stralingsfysica, Celestijnenlaan 200D, 3001 Leuven, Belgium}
\author{U. K\"oster}
\affiliation{Institut Laue Langevin, 71 Avenue des Martyrs, 38042 Grenoble, France}
\author{S. Kraemer}
\affiliation{KU Leuven, Instituut voor Kern- en Stralingsfysica, Celestijnenlaan 200D, 3001 Leuven, Belgium}
\affiliation{Ludwig-Maximilians-Universit\"at M\"unchen, Am Coulombwall 1, 85748 Garching, Germany}
\author{M. Laatiaoui}
\affiliation{Department Chemie - Standort TRIGA, Johannes Gutenberg-Universität Mainz, D-55099 Mainz, Germany}
\author{R. Lica}
\affiliation{Horia Hulubei National Institute of Physics and Nuclear Engineering, Bucharest, Romania}
\author{C. Merckling}
\affiliation{Imec, Kapeldreef 75, Leuven B-3001, Belgium}
\author{I. Morawetz}
\affiliation{Vienna Center for Quantum Science and Technology, Atominstitut, TU Wien, 1020 Vienna, Austria}
\author{H.W.T. Morgan}
\affiliation{Department of Chemistry and Biochemistry, University of California, Los Angeles, CA 90095, USA}
\author{D. Moritz}
\affiliation{Ludwig-Maximilians-Universit\"at M\"unchen, Am Coulombwall 1, 85748 Garching, Germany}
\author{L.M.C. Pereira}
\affiliation{KU Leuven, Quantum Solid State Physics, 3001 Leuven, Belgium}
\author{S. Raeder}
\affiliation{GSI Helmholtzzentrum für Schwerionenforschung GmbH, D-64291 Darmstadt, Germany}
\author{S. Rothe}
\affiliation{CERN SY-STI, Espl. des Particules 1, Geneva, 1211, Switzerland}
\author{F. Schaden}
\affiliation{Vienna Center for Quantum Science and Technology, Atominstitut, TU Wien, 1020 Vienna, Austria}
\author{K. Scharl}
\affiliation{Ludwig-Maximilians-Universit\"at M\"unchen, Am Coulombwall 1, 85748 Garching, Germany}
\author{T. Schumm}
\affiliation{Vienna Center for Quantum Science and Technology, Atominstitut, TU Wien, 1020 Vienna, Austria}
\author{S. Stegemann}
\affiliation{CERN SY-STI, Espl. des Particules 1, Geneva, 1211, Switzerland}
\author{J. Terhune}
\affiliation{Department of Physics and Astronomy, University of California, Los Angeles, CA 90095, USA}
\author{P.G. Thirolf}
\affiliation{Ludwig-Maximilians-Universit\"at M\"unchen, Am Coulombwall 1, 85748 Garching, Germany}
\author{S.M. Tunhuma}
\affiliation{KU Leuven, Quantum Solid State Physics, 3001 Leuven, Belgium}
\author{P. Van Den Bergh}
\affiliation{KU Leuven, Instituut voor Kern- en Stralingsfysica, Celestijnenlaan 200D, 3001 Leuven, Belgium}
\author{P. Van Duppen}
\affiliation{KU Leuven, Instituut voor Kern- en Stralingsfysica, Celestijnenlaan 200D, 3001 Leuven, Belgium}
\author{A. Vantomme}
\affiliation{KU Leuven, Quantum Solid State Physics, 3001 Leuven, Belgium}
\author{U. Wahl}
\affiliation{Centro de Ciências e Tecnologias Nucleares, Instituto Superior Técnico, Universidade de Lisboa, 2695-066 Bobadela LRS, Portugal}
\author{Z. Yue}
\affiliation{School of Physics, Engineering and Technology, University of York, York, YO10 5DD, United Kingdom}

\date{\today}

\begin{abstract}
A comparative vacuum ultraviolet spectroscopy study conducted at ISOLDE-CERN of the radiative decay of the $^{229m}$Th nuclear clock isomer embedded in different host materials is reported. The ratio of the number of radiative decay photons and the number of $^{229m}$Th embedded are determined for single crystalline CaF$_2$, MgF$_2$, LiSrAlF$_6$, AlN, and amorphous SiO$_2$. For the latter two materials, no radiative decay signal was observed and an upper limit of the ratio is reported. The radiative decay wavelength was determined in LiSrAlF$_6$ and CaF$_2$, reducing its uncertainty by a factor of 2.5 relative to our previous measurement. This value is in agreement with the recently reported improved values from laser excitation.
\end{abstract}

\maketitle

\section{\label{sec:intro}Introduction}
The $^{229}$Th nucleus contains a unique low-lying isomeric state which is accessible to laser spectroscopy in the vacuum-ultraviolet (VUV) regime\,\cite{kroger-1976,wense-2016,kraemer-2023}. Its existence allows for the development of a nuclear clock operated with VUV lasers which exist today, possibly providing an improved stability when compared to atomic clocks\,\cite{Peik-2003,campbell-2012}. A nuclear clock also has the potential to widen the breadth of applications by touting ultralight dark matter detection\,\cite{arvanitaki-2015,wense-2018}, improved geodesy\,\cite{blewitt-2015,thirolf-2019} and global positioning system (GPS) accuracy\,\cite{ashby-2003}, and sensitivity to variations of fundamental constants\,\cite{litvinova-2009,rellergert-2010}.

After the detection of the internal electron conversion decay of the $^{229}$Th nuclear clock isomer was achieved\,\cite{wense-2016} and following the approach described in Ref.\,\cite{verlinde-2019}, the $^{229m}$Th radiative decay was observed in an experiment whereby the isomer was populated via the $\beta$-decay of $^{229}$Ac produced from its shorter-lived precursors implanted in CaF$_2$ and MgF$_2$ wide band gap crystals \cite{kraemer-2023}. The implanted nuclei decayed to $^{229}$Ac which in turn $\beta$-decayed to the $^{229m}$Th isomeric state with a branching ratio in the range of 14 to 93\%\,\cite{verlinde-2019}. The photons from the decay of the isomer were observed in a VUV spectrometer and a wavelength value of $148.7 \pm 0.4$\,nm was obtained, whose uncertainty expressed as a photon frequency is on the order of THz. This opens the route for the development of a solid-state based nuclear clock. Recently, based on\cite{kraemer-2023} and the present work, laser excitation of $^{229}$Th embedded in a CaF$_2$ crystal \cite{beeks-2023-crystals} and a LiSrAlF$_6$ crystal\,\cite{elwell-2024} was achieved resulting in an improved wavelength value whose uncertainty is on the order of GHz \cite{tiedau-2024,elwell-2024}. Subsequently, narrowband laser excitation with a VUV frequency comb of $^{229}$Th in CaF$_2$ has determined the radiative decay transition frequency on the order of a few hundred kHz, allowing to resolve the quadrupole splitting of the nuclear resonance\cite{zhang2024dawnnuclearclockfrequency}.

Using crystals with a band gap wider than the $^{229m}$Th radiative decay energy hinders the internal electron conversion channel of $^{229m}$Th, resulting in the emission of VUV photons.Alongside CaF$_2$, other large band gap crystals, for example LiSrAlF$_6$, are being considered as hosts for $^{229}$Th \cite{Jackson-2009}.

With the $^{229m}$Th laser excitation achieved, comparing different crystal materials represents an important step in the development of a (solid-state) nuclear clock. This manuscript reports the results from a measurement campaign using the same methodology reported in Ref.\,\cite{kraemer-2023} to test different crystal materials and determine the radiative decay wavelength value.

\section{\label{sec:exp}Experimental Setup}
The experiment took place at the Isotope Separator On-Line DEvice (ISOLDE) facility\,\cite{catherall-2017} at CERN. A beam of 1.4\,GeV protons at $2\,\mu$A was impinged upon a $\mathrm{ThC}_x$ target producing $^{229m}$Th precursors which diffused to the surface by heating the target to $2000\,\degree$C. An ion beam of surface-ionized $^{229}$Ra$^+$ and $^{229}$Fr$^+$ was accelerated to 30\,keV and separated according to its mass-to-charge ratio through the General Purpose mass Separator (GPS). The radioactive ion beam was subsequently transported to the VUV spectrometer (Resonance Ltd., VM180)\,\cite{kraemer-2022,kraemer-2023-vuvsetup,kraemer-2023}.

A schematic diagram of the experimental setup can be found in Refs.\,\cite{kraemer-2023,kraemer-2023-vuvsetup}. The radioactive beam was implanted into crystals (see Tab.\,\ref{tab:crystal-specs}) that were mounted on a rotatable wheel (see Refs.\,\cite{kraemer-2022,kraemer-2023-vuvsetup} for details). For the radiative decay fraction measurements the implantation times varied between 10 and 30 minutes, however, wavelength determination measurements had longer implantation times between 40--90 minutes. After implantation, the wheel was rotated by $180\degree$ to place the crystals in front of the adjustable entrance slit of the spectrometer. A stepper motor controlled the rotation of the crystal holder to improve the reproducibility of crystal position compared to the manual rotation method used in the previous campaign reported in  Ref.\,\cite{kraemer-2023}. A deviation in the relative source position in front of the entrance slit based on the full step angle and the step angle accuracy of the motor driven wheel, given by the manufacturer, is negligible. VUV photons from the radiative decay entered the spectrometer, were then reflected to a diffraction grating in a Czerny-Turner configuration, and detected with a photomultiplier tube (PMT). A 250\,$\mu$m slit size was used for wavelength determination measurements and a 2\,mm slit size for efficiency and half-life measurements. In addition, a microchannel plate (MCP) detector with a phosphor screen and camera were installed at the radioactive beam position behind the wheel to perform beam diagnostics and identify the implantation location and size of the radioactive beam on the crystals.

For the wavelength measurements, the thin film CaF$_2$ crystal (CaF$_2$ 350 in Tab. \ref{tab:crystal-specs}) was used to reduce the Cherenkov background resulting from beta decays of implanted isotopes in the sample\,\cite{kraemer-2023}. Wavelength calibration measurements from a VUV light source consisting of a plasma discharge in air were taken before and after each series of wavelength measurements to account for any fluctuations in centroid position of the calibration lines.

In order to determine the number of implanted $^{229}$Fr and $^{229}$Ra isotopes, a high-purity germanium (HPGe) coaxial detector (Canberra, 75\% efficiency) with an aluminum cap was placed 1\,m away from the implantation point. Energy and absolute efficiency calibration of the HPGe detector were performed with activity calibrated $\gamma$-ray sources of $^{60}$Co, $^{137}$Cs, $^{152}$Eu, and $^{207}$Bi.

\subsection{Beam Diagnostics\label{sec:beamdiag}}
\begin{figure}[h]
    \centering
    \includegraphics[width=0.48\textwidth]{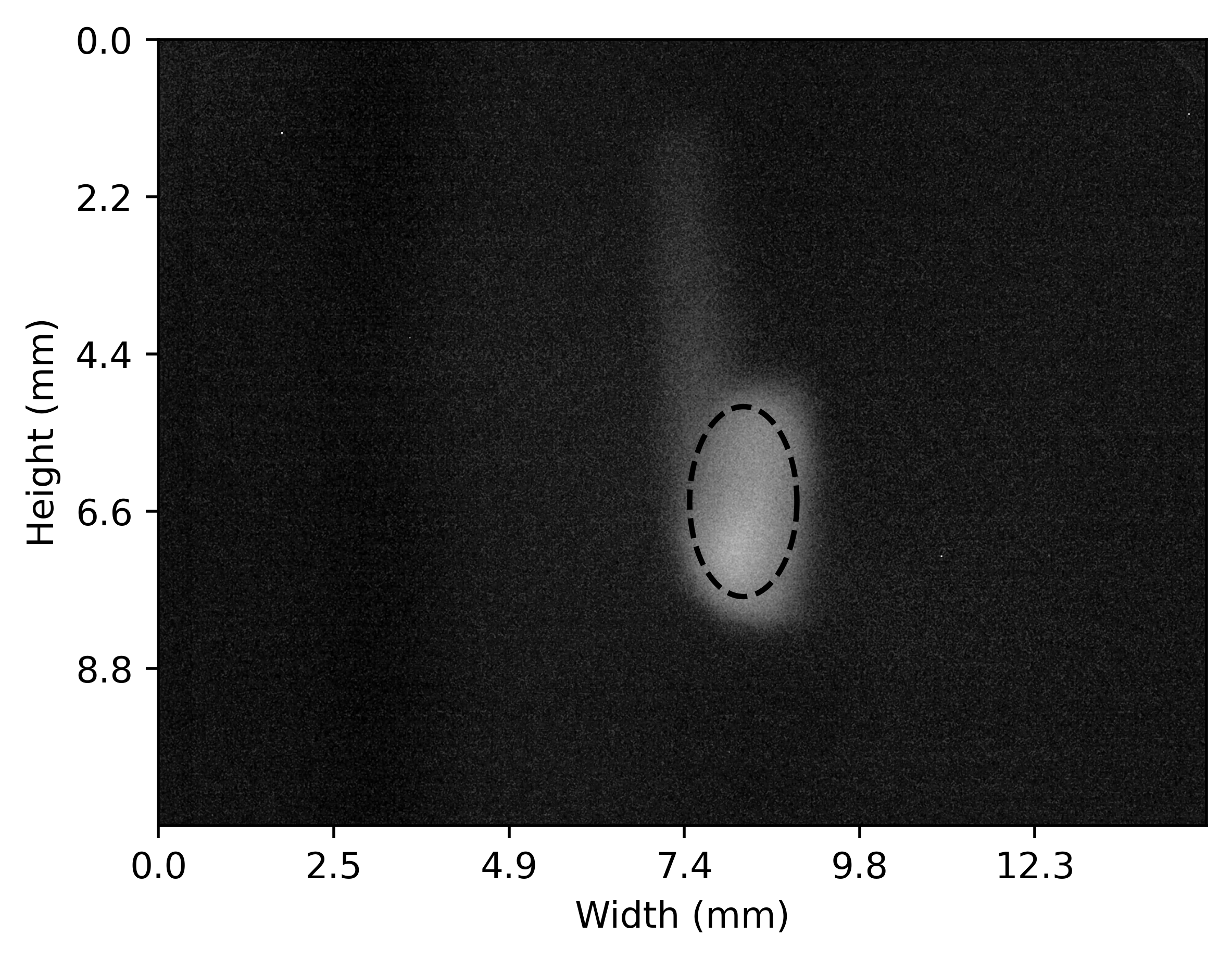}
    \caption{Image of the beam spot captured from the MCP+phosphor screen setup captured by a CCD camera. The dashed ellipse indicates the beam size at the full width at half maximum level. The dimensions of this ellipse are $1.5\,\mathrm{mm}\times 2.7$\,mm. The $x$ and $y$ axes correspond to the horizontal and vertical beam position, respectively. Axis tick labels are rounded to the nearest decimal in mm.}
    \label{fig:mcp}
\end{figure}

\begin{figure*}[htp]
    \centering
    \includegraphics[width=\textwidth]{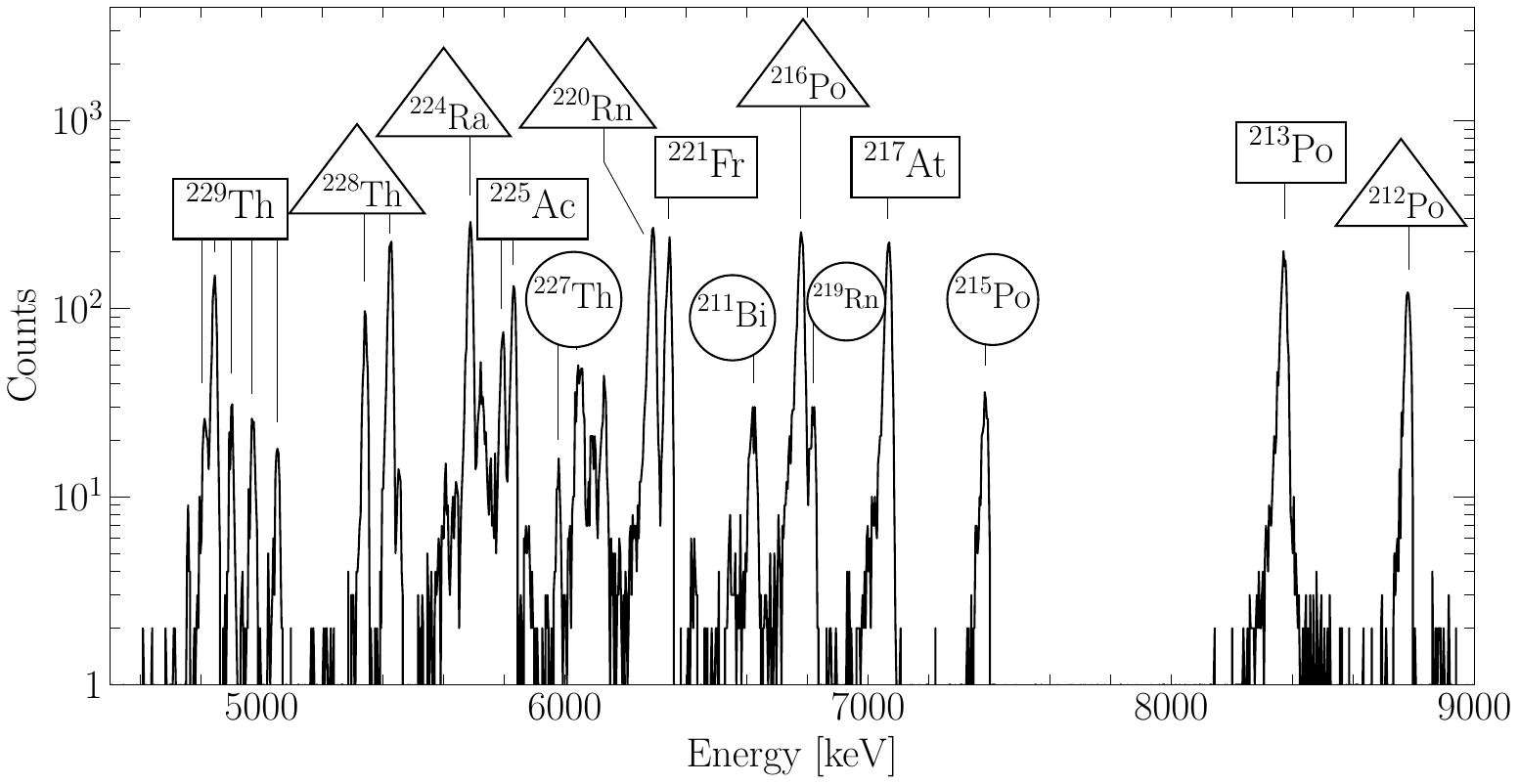}
    \caption{The $\alpha$-decay spectrum of the CaF$_2$ 350 sample (see Tab. \ref{tab:crystal-specs} for details of the crystal), the $\alpha$-lines from the $A=229$ (square), $A=228$ (triangle), and $A=227$ (circle) decay chains are indicated.}
    \label{fig:alpha_spec}
\end{figure*}

The radioactive ion beam was directed through an empty slot on the wheel crystal holder and impinged on the MCP detector that was equipped with a phosphorous screen anode. The screen was observed with a camera and images of the ion beam were collected. An example is shown in Fig.\,\ref{fig:mcp}. The intensity of the image was projected onto the $x$ and $y$ axes and fitted with a Gaussian profile to determine the full width at half maximum of the beam spot for each axis. The axes were calibrated using an optically visible image of equally spaced grooves, leading to a conversion factor of 87\,pixels/mm. The ion beam had a size of approximately $1.5\,\mathrm{mm}\times 2.7$\,mm at the full width at half-maximum.

\subsection{\label{sec:dec-spec}Radioactive Beam Intensities and Purity}
To quantify radiative decay fractions on different samples, the implanted amount of $^{229}$Fr ($t_{1/2}$\;=\;$50.2\pm2.0$\;s), $^{229}$Ra ($t_{1/2}$\;=\;$4.0\pm0.2$\;min), and $^{229}$Ac ($t_{1/2}$\;=\;$62.7\pm0.5$\;min) isotopes were quantified. These amounts depend on the implantation time and the beam intensity of the specific isotopes. Using Bateman's equations with the known half-lives and $\beta$-decay branching ratio range of $^{229}$Ac feeding $^{229m}$Th, the amount of $^{229}$Th isomers can be determined as a function of time.

The beam intensities of $^{229}$Fr and $^{229}$Ra were determined using their characteristic $\gamma$-ray transitions measured with the HPGe detector during implantation. Based on the time behavior of the intensity of the characteristic $\gamma$ transition from the $^{229}$Ac $\beta$-decay, it was concluded that the beam intensity of $^{229}$Ac $\mathcal{O}(10^6$\,pps) was negligible compared to $^{229}$Ra.
Information on the $\beta$-decay scheme of $^{229}$Ra is limited as no absolute $\gamma$-ray intensities are reported \cite{NDS229Ra}. To improve this, the $A=229$ beam was directed to the ISOLDE Decay Station (IDS) \cite{IDS}, and several intense $\gamma$-lines could be attributed to the decay of $^{229}$Ra feeding levels in $^{229}$Ac. By comparing the observed decays from the newly identified $\gamma$ transitions with the 569\,keV $\gamma$-line of $^{229}$Ac, for which the absolute intensity is reported \cite{NDS229Ac}, the absolute $\gamma$-ray intensities of the $^{229}$Ra lines were determined. The results of this decay study will be reported in a forthcoming publication \cite{elskens-2024}. The intensity of the characteristic $\gamma$-rays of $^{229}$Fr, $^{229}$Ra and $^{229}$Ac was measured as a function of time during implantation. The time evolution was fitted with Bateman's equations and after correction for $\gamma$-ray efficiency and absolute intensity, the beam intensity was determined.

To verify the procedure and to determine precisely the beam purity, $\alpha$ spectroscopy was performed on the implanted crystals about 10 months after the experiment. A $10\,\mathrm{mm}\times10$\,mm Si PIN photodiode detector (Hamamatsu S3590-09) was used with a collimator which limited the detection area to $3.25\,\mathrm{mm}\times3.25$\,mm to avoid edge effects from the surface of the detector. The $\alpha$-decay setup was simulated to determine the detection efficiency. This was further verified with measurements taken at different detector-to-crystal distances using an intensity calibrated $^{241}$Am source. To reduce the uncertainty on the $\alpha$ detection efficiency, the final measurements were performed with a detector-to-crystal distance of 6\,cm. The crystals were all measured at this distance for several weeks to optimize the statistics. A typical $\alpha$-decay energy spectrum is shown in Fig. \ref{fig:alpha_spec}. These spectra also allowed to determine the beam composition. For the $A/Q=229$ radioactive ion beam setting of the mass separator this resulted in: $A=229$ is $99.47\pm0.04$\%, $A=228$ is $0.49\pm0.04$\%, and $A=227$ is $0.04\pm0.01$\%. Contaminants of neighboring masses are due to the finite mass resolution and tailing from neighboring mass peaks. From the $\alpha$-lines from $^{229}$Th \cite{NDS229Th}, it was possible to calculate the implanted activity and estimate an average implantation rate for each crystal. In addition to the $\alpha$-decay setup the implanted crystals were measured with an intensity calibrated HPGe detector, and the $\gamma$-lines from $^{229}$Th and their absolute $\gamma$-ray intensities \cite{NDS229Th} were used to estimate the implanted activity. This was found to be in good agreement with the value obtained from the $\alpha$-decay measurements. However, the total implantation activity obtained from the off-line $\alpha$ and $\gamma$ spectroscopy measurements did not agree with the values deduced from the on-line $\gamma$ spectroscopy measurements suggesting that the literature value \cite{229Ac_gamma_spec_ref} of the absolute $\gamma$-ray intensity of the $^{229}$Ac decay line used to estimate the absolute intensities of the $^{229}$Ra $\gamma$-lines needs to be corrected. The implanted activity determined from the offline measurements was about a factor of 2 smaller than the one deduced from the online implantations using the observed $\gamma$-ray intensities.  Therefore, a correction factor of $0.53\pm0.03$ to the rates obtained from the online $\gamma$ decay measurements was implemented, and the final average values for the beam intensities were found to be $1.42\pm0.09\cdot10^6$\;pps for $^{229}$Fr, and $1.24\pm0.08\cdot10^8$\;pps for $^{229}$Ra. These rates were used to estimate the radiative decay fraction.

\subsection{Crystal Specifications}
\begin{table}[th]
\caption{\label{tab:crystal-specs}
Specifications on each crystal used in the experiment. Amorphous SiO$_2$ is denoted as a-SiO$_2$ in this table. The Si(111) substrate was 0.75\,mm thick, whereas the Si(100) substrate was 0.67\,mm thick. The thickest of the crystals (5\,mm thickness) are denoted `bulk'.}
\begin{ruledtabular}
\begin{tabular}{cccccc}
Crystal & Dimensions ($l\times w \times t$) & Substrate & Manufacturer \\
\hline
CaF$_2$ 350 & $2\,\mathrm{cm}\times 2\,\mathrm{cm} \times 50\,\mathrm{nm}$ & Si(111) & Imec   \\
CaF$_2$ 850 & $2\,\mathrm{cm}\times2\,\mathrm{cm}\times 50\,\mathrm{nm}$ & Si(111) & Imec  \\
AlN & $1\,\mathrm{cm}\times 1\,\mathrm{cm} \times 50\,\mathrm{nm}$ & Si(111) & Imec  \\
a-SiO$_2$ & $2\,\mathrm{cm}\times 2\,\mathrm{cm} \times 300\,\mathrm{nm}$ & Si(100)  & Imec  \\
LiSrAlF$_6$& $1\,\mathrm{cm}\times 1\,\mathrm{cm} \times 1\,\mathrm{mm}$ & --  & AC Materials  \\
MgF$_2$ bulk & $\varnothing=25.4\,\mathrm{mm} \times 5\,\mathrm{mm}$ & -- & Thorlabs, Inc. \\
CaF$_2$ bulk & $\varnothing=25.4\,\mathrm{mm} \times 5\,\mathrm{mm}$ & -- & Thorlabs, Inc. \\
\end{tabular}
\end{ruledtabular}
\end{table}

A summary of the specifications of the different crystals used in this experiment is shown in Tab.\,\ref{tab:crystal-specs} and the band gaps for each material are tabulated in Tab.\,\ref{tab:bandgaps}. The CaF$_2$ epitaxial thin-films were grown using molecular beam epitaxy (MBE) at Imec (Leuven, Belgium), using the same growth parameters except for different growth temperatures, indicated by the number in degrees Celsius in the first column of Tab.\,\ref{tab:crystal-specs}, which is expected to result in different lattice defect densities and stacking sequence due to distinct Si/CaF$_2$ interfaces\cite{wang2002epitaxy}. MBE is a process used to grow thin layers of materials with precise interface, thickness, and uniformity under ultra high vacuum conditions \cite{FRANCHI-2013}. Before being placed in the MBE reactor, the substrates, 0.75\,mm thick wafers of Si(111), were cleaned with 2\% NH$_4$F for 60 seconds followed by high thermal annealing up to 875\degree C in ultra-high vacuum to remove oxides and organic contaminants. Once inside the growth chamber the 50\,nm thin-films were grown epitaxially at temperatures of 350\degree C and 850\degree C, leading to A-type stacking (F-interface) and B-type stacking (CaF interface), respectively.
More details on the growth procedure can be found in Ref.\,\cite{kraemer-2023}. The AlN epitaxial thin film was grown at Imec using metalorganic chemical vapour deposition (MOCVD) \cite{chang-2020}. The amorphous SiO$_2$ (a-SiO$_2$) thin film on Si was prepared at Imec and consists of a dry thermal oxide layer created by annealing a Si wafer in a controlled O$_2$ environment, as widely used in microelectronics.

\begin{table}[h]
\caption{\label{tab:bandgaps}
Band gaps for each host material. A host material with a band gap larger than the radiative decay energy of $^{229}$Th ($8.4\,\mathrm{eV}$ \cite{tiedau-2024,elwell-2024}) is expected to be a necessary condition to suppress the internal conversion decay channel \cite{kraemer-2023}. The band gap of amorphous SiO$_2$ (or a-SiO$_2$), which is less well-defined than for the other materials in this list (crystalline), is discussed in the text.}
\begin{ruledtabular}
\begin{tabular}{ccc}
Material & Band gap (eV) & Band gap Ref.  \\
\hline
MgF$_2$ &  12.4 & \cite{thomas-1973} 
\\
CaF$_2$ & 12.1 & \cite{rubloff-1972}\\
LiSrAlF$_6$ & 12.1 & \cite{shiran-2005} \\
a-SiO$_2$ & 9.3-9.6 & \cite{weinberg1979transmission, ravindra-1986} 
\\
AlN & 6.2 & \cite{edgar-1990} \\
\end{tabular}
\end{ruledtabular}
\end{table}

The LiSrAlF$_6$ crystal was produced by AC Materials for University of California, Los Angeles (UCLA) using the Czochralski method \cite{klimm-2000}. The bulk crystals, CaF$_2$ (crystal identifier nr. WG51050) and MgF$_2$ (WG61050) broadband uncoated precision windows, were purchased from Thorlabs, Inc. All thin-film crystals, as well as the LiSrAlF$_6$, were mounted on near-IR absorptive neutral density filters (Thorlabs, Inc., $\varnothing=25\,\mathrm{mm} \times 2.8\,\mathrm{mm}$, OD:\,6.0).

\section{Radiative Decay Spectra}
\begin{figure}[htp]
    \centering
    \includegraphics[width=0.48\textwidth]{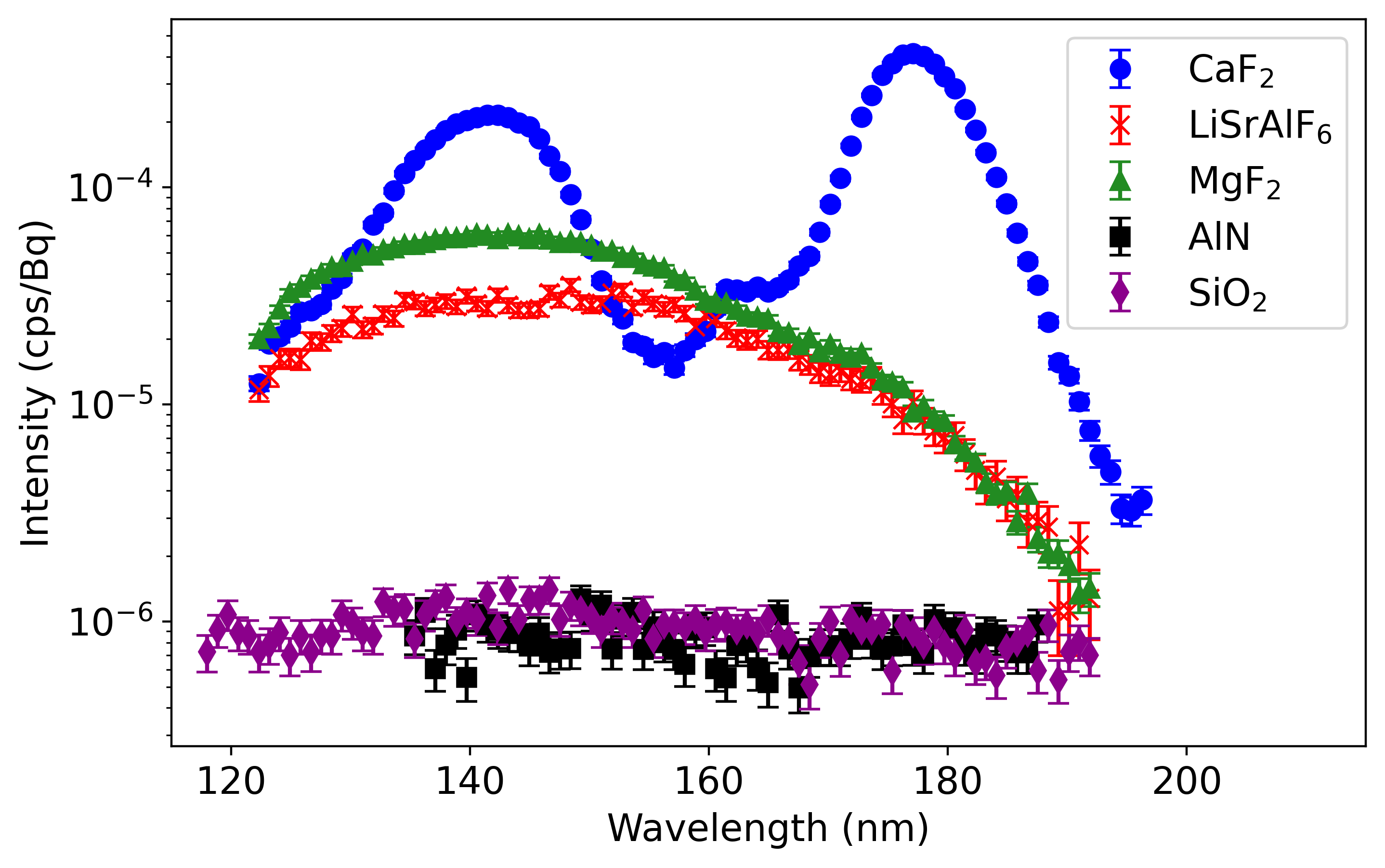}
    \caption{(Color online) Cherenkov spectra from $A=230$ implantations for CaF$_2$ (circles), MgF$_2$ (triangles), LiSrAlF$_6$ (x's), AlN (squares) and SiO$_2$ (diamonds) normalized by the activity deduced from the $\gamma$-data recording during implantation. Broad radioluminescence peaks are present in CaF$_2$ bulk at approximately 142\,nm and 177\,nm. All spectra were taken with a measurement time of 10 seconds per data point and a step size of 100 motor positions between each point.}
    \label{fig:cherenkov-spectra}
\end{figure}

\begin{figure*}[ht]
    \centering
    \includegraphics[width=\textwidth]{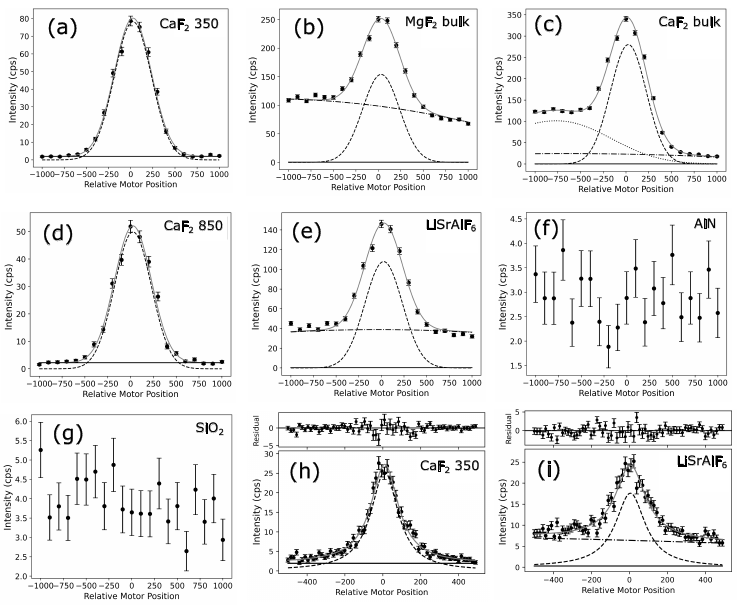}
    \caption{Typical spectra taken with a 2\,mm (a-g) and 250\,$\mu$m (h,i) entrance slit demonstrating the fitting method for the radiative decay of $^{229m}$Th. Measurements (a-g) were taken with a measurement time of 10 seconds per point and a step size of 100 motor positions between each point. Measurements (h,i) were taken with a 10 second measurement time per point and a step size of 15 motor positions between each point. The dashed curve is a Gaussian peak in (a to e) and a Lorentzian peak in (h and i). The  dot-dashed curve is a polynomial representing the Cherenkov background radiation and the dotted curve is the radioluminescence peak possibly due to a color center. The solid black line is the dark count rate from the PMT. The reduced Cherenkov background in case of spectra taken with the thin film crystals (a, d, h) did not allow to distinguish clearly between the Cherenkov background from the dark countrate. (a) CaF$_2$ 350 thin film, (b) MgF$_2$ bulk, (c) CaF$_2$ bulk, (d) CaF$_2$ 850, (e) LiSrAlF$_6$, (f) AlN, and (g) SiO$_2$, (h) CaF$_2$ 350 and (i) LiSrAlF$_6$.}
    \label{fig:all-spectra}
\end{figure*}

The methodology used to fit the radiative decay spectra was adapted from Refs.\,\cite{kraemer-2022,kraemer-2023}. Spectra dedicated to the determination of radiative decay fraction were acquired when the maximum number of $^{229m}$Th were present based on results from the Bateman equations (see section \ref{sec:dec-spec}). Spectra obtained from $A=230$ implantations are shown in Fig.\,\ref{fig:cherenkov-spectra} while sample spectra from $A=229$ are shown in Fig.\,\ref{fig:all-spectra}(a-g).

The $A=229$ spectra obtained with a 2 mm slit size were fitted using three components: a constant PMT dark count rate, the Cherenkov background and the isomeric VUV signal. These are shown separately in Fig. \ref{fig:all-spectra}. The Cherenkov background was modeled by a second-degree polynomial where the coefficients were constrained by the Cherenkov radiation observed from an implantation of $A=230$ into each respective crystal (Fig.\,\ref{fig:cherenkov-spectra}). The VUV signal was described by a  Gaussian function. 

A modification to this fitting procedure was applied to data obtained with the 5\,mm-thick CaF$_2$ bulk crystal, which showed a broad radioluminescence peak excited by the Cherenkov radiation and emitting at approximately 142\,nm on the left-hand side of the $^{229m}$Th peak (see Fig.,\ref{fig:cherenkov-spectra}). An asymmetric Gaussian profile was used in addition to the polynomial and its parameters were constrained to fit results obtained from Fig.\,\ref{fig:cherenkov-spectra}.

For the wavelength determination measurements with a $250\,\mu$m slit size spectra were obtained when the $\beta$-decay of $^{229}$Ac and $^{229m}$Th were in transient equilibrium. The fitting procedure uses the same approach including a decay correction detailed in Ref.\,\cite{kraemer-2022} and a Lorentzian profile  instead of a Gaussian, since the Lorentzian describes the peak better at the smaller slit size. Spectra for both the CaF$_2$ 350 thin film and the LiSrAlF$_6$ are shown in Fig.\,\ref{fig:all-spectra}(h,i).

\section{\label{sec:vuv-eff}Radiative Decay Fraction} 
The radiative decay fraction of a crystal is proportional to the ratio between the number of $^{229m}$Th decaying via radiative decay versus the number of $^{229m}$Th embedded in the crystal at the time of the measurement. This efficiency is defined in Eq.\,\ref{eq:efficiency}, where $H$ is the signal height determined from the radiative decay spectrum measured at a point in time closest to when the number of $^{229m}$Th was at its maximum, $N_{229m}$ is the number of $^{229m}$Th nuclei present at the time when the radiative decay spectrum was measured and is dependent on the half-life of the isomer, and $\varepsilon_I$ is the total detection efficiency of the instrument. The decay constant is $\lambda_{229m}=\ln(2)/t_{1/2}$, where $t_{1/2}$ is the half-life of the $^{229m}$Th isomer deduced from the time behaviour of the VUV signal in the different crystals.

The efficiency of the instrument is $\varepsilon_I=\varepsilon_{P}\varepsilon_{G}\varepsilon_{C}$, where $\varepsilon_{P}=19\%$ is the quantum efficiency of the PMT detector and $\varepsilon_\mathrm{G}=40\%$ is the grating efficiency, both specified by the manufacturer. The $\varepsilon_\mathrm{C}$ is the collection efficiency of the VUV photons based on the solid angle of the entrance slit and the collimation mirror for the emitted radiation from the source \cite{kraemer-2023-vuvsetup}. The collection efficiency was simulated for each crystal according to their different thicknesses and consequently their distance relative to the entrance slit (see Tab.\,\ref{tab:crystal-specs}). A collection efficiency of 2.4\% was determined for CaF$_2$ bulk and used to calculate a lower limit for its absolute radiative decay fraction.

The VUV transmission through the crystal is assumed to be 100\% due to the small implantation depth of 17\,nm in CaF$_2$ and reflection at the crystal surfaces is assumed to be negligible . Also, the reflectivity from both the collimation mirror and the focusing mirror of the spectrometer are assumed to be 100\% \cite{kraemer-2023}. The VUV-efficiency thus becomes
\begin{equation}
    \label{eq:efficiency}
    \varepsilon_{\mathrm{VUV}}=\frac{H}{\lambda_{229m}\,N_{229m}(\lambda_{229m})\,\varepsilon_I} .
\end{equation}

As the efficiencies of the instrument and the VUV transmission through the crystal should be considered as upper limits, only a lower limit of the absolute radiative decay fraction can be reported. For CaF$_2$, a lower limit of the absolute radiative decay fraction of 0.9\% or 5.7\% was obtained assuming the 93\% or 14\% $^{229}$Ac $\beta$ branching ratio to the $^{229m}$Th, respectively.

\begin{figure}[ht]
    \centering
    \includegraphics[width=0.48\textwidth]{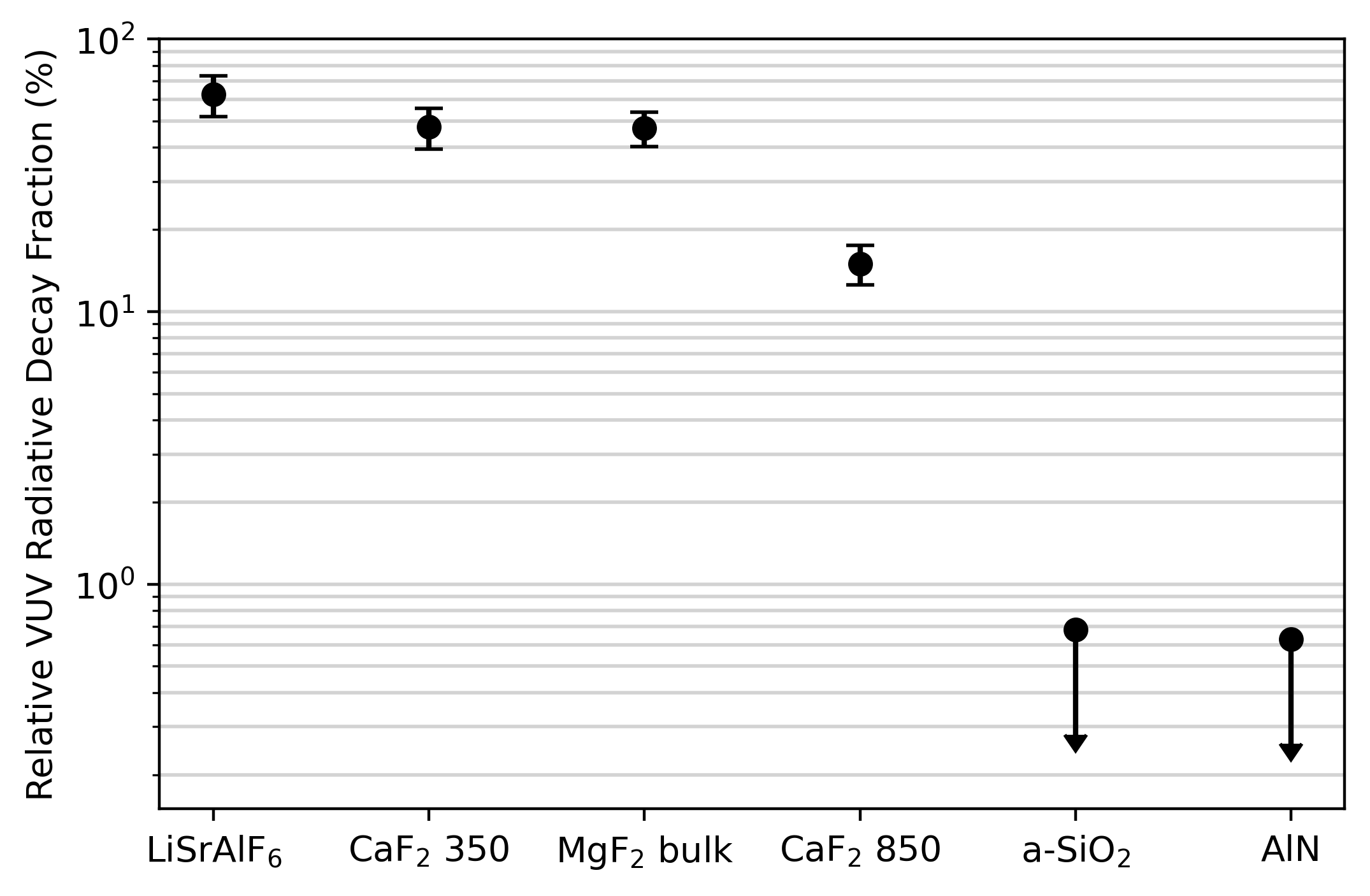}
    \caption{Radiative decay fractions for each crystal relative to CaF$_2$ bulk. An upper limit is reported for AlN and SiO$_2$ due to the absence of a signal, therefore a downward-pointing arrow is shown for both of these crystals.}
    \label{fig:relative-efficiencies-plot}
\end{figure}

The relative radiative decay fraction between the different host materials are shown in Tab.\,\ref{tab:vuv-efficiency} and Fig.\,\ref{fig:relative-efficiencies-plot}. Uncertainties for the decay rates from each implantation were propagated through Bateman's equations and a weighted average radiative decay fraction for each crystal was obtained (the result from Eq.\,\ref{eq:efficiency}). The value obtained for the crystal with the highest efficiency, CaF$_2$ bulk, was used to normalize each crystal efficiency to obtain the relative efficiencies listed in Tab.\,\ref{tab:vuv-efficiency}. To cope with the varying half-life of the $^{229m}$Th in different crystal materials, radiative decay fractions were calculated by varying the half-life between $t_{1/2}=500$ and $t_{1/2}=800$ seconds which covers the range of half lives in the different crystals \cite{kraemer-2023,elskens-2024}. The differences in the radiative decay fraction results using these two extreme values were considered as a systematic uncertainty. Additionally, the relative difference between the maximum number of $^{229m}$Th nuclei according to Bateman's equation and the number of $^{229m}$Th nuclei at the time of the radiative decay measurement is on the order of 1\% and is also considered as a systematic uncertainty.  The statistical and systematic uncertainties were added in quadrature and reported in Tab.\,\ref{tab:vuv-efficiency}. 

\begin{table}[th]
\caption{\label{tab:vuv-efficiency}
Relative radiative decay fractions for each crystal. The total number of implantations $N_\mathrm{tot}$ and total implantation time $t_\mathrm{tot}$ are also reported. All crystals, except for the thin film CaF$_2$ 350 and 850, received one $A=230$ implantation. Note that the CaF$_2$ 350 crystal was used for wavelength measurements resulting in a large total implantation time.}
\begin{ruledtabular}
\begin{tabular}{cccc}
Crystal & Relative Decay Fraction (\%)  & $N_\mathrm{tot}$ & $t_\mathrm{tot}$ (s) \\
\hline
CaF$_2$ bulk & 100 & 5 & 6523\\
LiSrAlF$_6$& $62.7\pm10.7$ & 3 & 5517 \\
CaF$_2$ 350 & $47.5\pm8.0$ & 7 & 16450\\
MgF$_2$ bulk & $47.1\pm6.9$ & 5 & 5525 \\
CaF$_2$ 850 & $15.0\pm2.5$ & 2 & 2094\\
a-SiO$_2$ & $<0.68$ & 2 & 2732 \\
AlN & $<0.63$ & 2 & 2602\\
\end{tabular}
\end{ruledtabular}
\end{table}

\subsection{\label{subsec:caf2,mgf2,lisaf} CaF$_2$, MgF$_2$ and LiSrAlF$_6$}

It is generally accepted that decay of the isomer via electron conversion can be suppressed if the Th atoms are incorporated into a crystal in a way that ensures that the gap between the highest-energy filled electronic state and the lowest-energy empty state remains larger than the isomer energy \cite{Dessovic2014,Pimon2020,Jackson2009,Tkalya2000,Ellis2014,Rellergert2010}. Within the $^{229}$Th community, this condition is often referred to as \emph{preserving the band gap}. In the present work, although the measured radiative decay fractions are likely affected by various material parameters and physical processes that are currently unknown, it is worthwhile discussing potential reasons for the differences observed among crystals. 

The highest efficiency was observed for the CaF$_2$ single crystal, indicating that a substantial fraction of the Th atoms take up configurations that suppress internal conversion, \emph{i.e.} substitute for Ca (as reported in \cite{kraemer-2023}) and likely form complexes with neighboring defects (\emph{e.g.,} with two F interstitials or with one Ca vacancy) yielding an effective 4+ charge state and a band gap larger than the isomer transition energy (\emph{preserving the band gap}) \cite{Dessovic2014}. Interestingly, the CaF$_2$ thin films show significantly lower efficiencies, and by different amounts, depending in this case on the film growth temperature. Compared to high-quality single crystals, the thin films are expected to have a higher density of defects, both point defects (because the MBE growth occurs far from thermodynamic equilibrium) and extended defects (e.g., grain boundaries, due to the lattice mismatch between the film and the substrate). Many of these defects may produce in-gap states (\emph{not preserving the band gap}), so that increasing the concentration of those defects (e.g., by growing a film at different temperatures, as here) increases the probability of such electronic states interacting with $^{229m}$Th nuclei and open electron conversion channels. Moreover, such defects may form complexes with the Th atoms, resulting in local Th configurations that themselves have in-gap states. The CaF$_2$ layer grown at 350\degree C with A-type stacking sequence is fully strained to the Si(111) substrate leading to a reduced defect density with respect to the CaF$_2$ layer gown at 850\degree C with B-type stacking sequence which is fully relaxed and therefore present a larger defectivity\cite{wang2002epitaxy}.  

Comparing our results on CaF$_2$, MgF$_2$ and LiSrAlF$_6$ crystals is even more challenging, because it involves not only different compounds but likely also significant variations in the types and concentrations of defects that may contribute to conversion channels. Nevertheless, simply looking at the available literature, one may argue that MgF$_2$ and LiSrAlF$_6$ would indeed be more prone to electron conversion channels, because in both cases the lowest-energy Th configurations fail to \emph{preserve the band gap} \cite{Pimon2020, elwell-2024}. In MgF$_2$, the lowest-energy configuration consists of a complex containing a Mg-substitutional Th and one Mg vacancy, which introduces in-gap states, resulting in a predicted band gap that is smaller than the isomer transition energy \cite{Pimon2020}. A configuration involving two F interstitials is predicted to preserve a band gap larger than the isomer energy, but is itself less energetically favorable (by about 0.8 eV). In LiSrAlF$_6$, the lowest-energy configuration consists of Th substituting for Sr with two neighboring F interstitials, while the next most favorable configurations (at a cost of about 0.6 eV) consists of Th in a Sr site next to one F interstitial and one Li vacancy; both configurations fail to preserve the band gap \cite{elwell-2024}. In the present work, the Th atoms are introduced by ion implantation, far from equilibrium, which appears to allow for the creation of Th configurations that preserve the band gap despite being energetically unfavorable (such as that involving the Mg vacancy in MgF$_2$), so that radiative decay is still observed. Nevertheless, it is plausible that in MgF$_2$ and LiSrAlF$_6$ larger fractions (compared to CaF$_2$) of the Th atoms take up configurations (such as the predicted lowest-energy ones) that allow for electron conversion, contrary to CaF$_2$, which would then explain why the radiative decay fraction is lower than for the CaF$_2$ crystal. Interestingly, this hypothesis would additionally suggest that, for MgF$_2$ and LiSrAlF$_6$, introducing the Th atoms by ion implantation may yield higher radiative decay fractions than doing so during growth, closer to equilibrium.

\subsection{AlN and SiO$_2$ Crystals}
Contrary to CaF$_2$, MgF$_2$ and LiSrAlF$_6$, no radiative decay signal was detected in either of the AlN or SiO$_2$ crystals (Fig.\,\ref{fig:all-spectra} (f,g)). The $A=229$ implantation measurements were fitted with a polynomial constrained to the Cherenkov spectra obtained from the $A=230$ implantation. If a signal were to be present, it was defined as being greater than or equal to a $3\sigma$ deviation from the residuals obtained by the $A=229$ fit leading to the upperbound on the relative radiative decay fraction reported in Tab.\,\ref{tab:vuv-efficiency}.

In the case of AlN, its band gap is lower than the $^{229m}$Th radiative decay energy, meaning the internal conversion channel is energetically favorable and a strong isomer decay signal was not expected. In contrast to AlN, amorphous SiO$_2$ has a band gap (9.3-9.6 eV \cite{weinberg1979transmission, ravindra-1986}) that is larger than the $^{229m}$Th transition energy. Although amorphous insulators have a non-zero electronic density of states (DOS) within the band gap, these states are typically more localized compared to valence or conduction band states. Such localized states can still, in principle, enable internal conversion channels. However, due to their localized nature, in order for such states to fully account for the complete suppression of the radiative decay fraction, they would have to be spatially correlated with the positions of the Th nuclei. For true band states (valence or conduction), such correlation is not required, because band states are fully delocalized across a crystal, but in SiO$_2$ the gap between valence and conduction bands is larger than the isomer transition energy, \emph{i.e.} electron conversion via such a channel is not allowed. 
Photon absorption is another potential reason for the non-observation of VUV emission from SiO$_2$. However, considering the absorption coefficient of dry amorphous SiO$_2$ (of the order of 1000 cm$^{-1}$ around 8.4 eV) \cite{vella2008vacuum}, one expects a transmission above 99\% through a 40 nm SiO$_2$ layer (which should contain more than 90\% of the emitting $^{229m}$Th nuclei). 
Therefore, the fact that the $^{229m}$Th radiative decay signal was not observed for SiO$_2$ strongly suggests that the introduction of Th atoms in  SiO$_2$ creates electronic states within the band gap. 
Detailed calculations on Th:SiO$_2$ addressing this question are expanded upon in the following section (Sec.\,\ref{sec:sio2-calc}).

\section{\label{sec:sio2-calc} Calculations of Thorium in a Silicon Dioxide Environment}

The electronic properties of thorium atoms in SiO$_2$ were modelled using density functional theory (DFT) calculations, using the supercell approach. The purpose of the calculations was to gain insight into in-gap electronic states potentially created by the Th atoms, in an attempt to understand why radiative decay was not observed for SiO$_2$. The model structures used in our DFT calculations are based on crystalline SiO$_2$ while the experiments use amorphous SiO$_2$. Nevertheless, the model still allows us to obtain relevant insight, since the main conclusions, as described below, are more related to local effects (localized Th states and the nature of Th-O bonds) rather than to the long-range order (that exists in our DFT model but not in the experiment).

\begin{figure}[ht]
    \centering
   \includegraphics[width=0.48\textwidth]{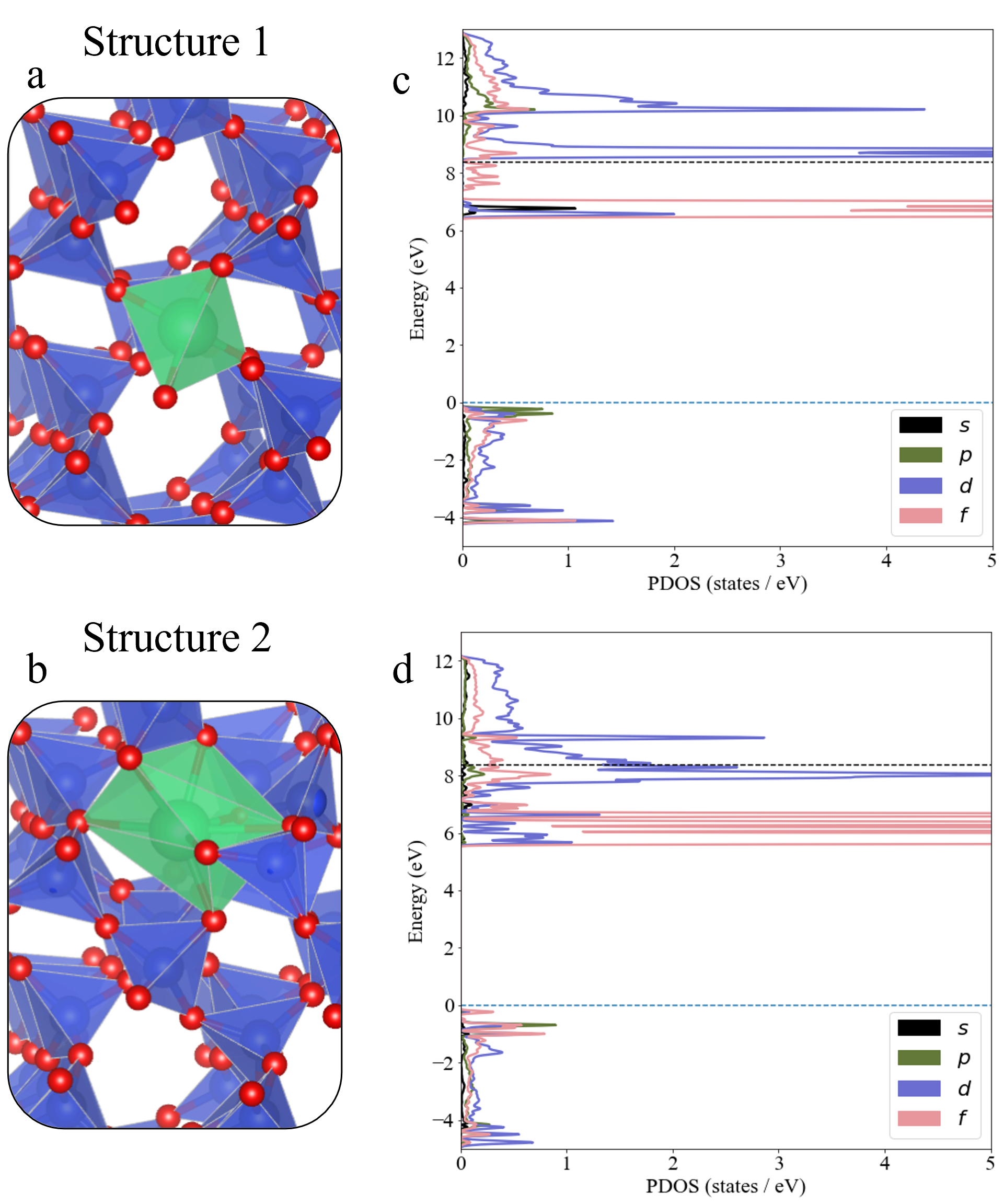}
   \caption{(a): DFT structure 1 for the Th defect in SiO$_2$. Th is shown in green, Si in blue, and O in red. (b): DFT structure 2 for the Th defect in SiO2. (c) Thorium projected density of states (PDOS) in Th:SiO2 defect structure 1. Energies are relative to the highest occupied band, and the black dashed line at 8.355 eV marks the $^{229}$Th isomer transition energy (note that since the nuclear excited state is not an electronic state, marking it relative to the Fermi energy is an arbitrary choice). (d) Thorium PDOS in structure 2.}
   \label{fig:sio2-calculations}
\end{figure}

Two structures were used to model Th:SiO$_2$, one where thorium replaces a silicon atom (structure 1) and one where thorium and two oxygen atoms are added to SiO$_2$ (structure 2). This is not an exhaustive exploration of all possible Th configurations in SiO$_2$, merely examples of likely configurations with a ThO$_4$ environment, which one expects to be more prone to a Th$^{4+}$ charge state and thus to suppress electron conversion. Naturally, many other structures are possible, in particular considering that the Th atoms were incorporated by ion implantation, far from equilibrium; however, the purpose of these calculations is to assess whether, even under the most favorable conditions for radiative decay that one can conceive, there are still characteristics specific to $\alpha$-SiO$_2$ that could create conversion channels. To generate these two structures, we started with the crystal structure of $\alpha$-SiO$_2$ (space group 152) and made a $3\times3\times3$ supercell with side lengths of approximately 15\,Å to minimize unphysical Th-Th interactions. In the first structure, one Si atom was replaced by Th (a substitutional defect), and in the second, a Th atom and two O atoms were added to voids in the SiO$_2$ lattice (an interstitial defect). Both structures were optimized. The resulting geometry for structure 1, shown in Fig.\,\ref{fig:sio2-calculations}, has Th in a ThO$_4$ tetrahedron that is significantly larger than the SiO$_4$ tetrahedra in the bulk material (5.12 Å$^3$ vs 2.21 Å$^3$) due to the larger size of Th$^{4+}$. In structure 2, Th is in a low-symmetry ThO$_7$ polyhedron. Formation energies for the defects were computed by writing balanced reactions between Th:SiO$_2$, pure SiO$_2$, and SiO$_2$. The formation energies are +2.4 eV for structure 1 and +7.9 eV for structure 2. Our intuition is that structure 2, featuring a higher coordination number for Th, is a better reflection of Th in amorphous SiO$_2$.  
The calculations were performed with Vienna Ab initio Simulation Package (VASP) \cite{RN12}, version 6.4.2, using the projector augmented-wave (PAW) \cite{RN14} method with a plane-wave cutoff of 500 eV and a spin-restricted formalism. The Perdew-Burke-Ernzerhof (PBE) \cite{RN13} functional was used for all structural optimizations, and the modified Becke-Johnson (MBJ) \cite{RN490,RN489} functional was used for density of states calculations. The unit cell of $\alpha$-SiO$_2$ was optimized with a 9-9-6 Gamma-centered k-point grid, and the thorium-doped $3\times3\times3$ supercell calculations were done with a 3-3-2 k-point grid (0.03 Å$^{-1}$ spacing in all cases). 
Polyhedral volumes were computed using visualization for electronic and structural analysis (VESTA) \cite{RN140,RN545}.

The Th orbital energies were determined through the projected density of states (PDOS) for each structure, shown in Figure \ref{fig:sio2-calculations}. These calculations used the MBJ functional, which has been previously used to study thorium defects in wide band gap crystals \cite{elwell-2024}. In both structures, empty Th states emerge within the band gap, and the energy difference between them and the top of the valence band is smaller than the isomer transition energy, so that internal conversion may be possible via excitation of an electron from oxygen (which dominates the top of the valence band) to these thorium-related in-gap states. The lowest unoccupied states are principally composed of Th 5$f$, with small admixture of Th 6$d$ and 7$s$, and O 2$p$ from the atoms closest to thorium. Higher unoccupied states also contain significant 5$f$ and 6$d$ character. In structure 2 the unoccupied bands are lower energy than in structure 1 and have a broader energy dispersion, which suggests that there are more pathways for internal conversion.

The Th PDOS in SiO$_2$ can be compared to that in Th-doped LiSrAlF$_6$ \cite{elwell-2024} in an attempt to understand why radiative nuclear decay has been observed in Th:LiSrAlF$_6$ but not Th:SiO$_2$, while in both cases in-gap states appear at energies lower than that of the isomer transition. The PDOS plots for SiO$_2$ show Th contributions to the valence band, which is principally composed of O 2$p$. Internal conversion would require the electronic wavefunction to overlap with the Th nucleus; a larger Th component in the valence band implies more overlap and thus higher conversion rates. This contribution is in fact consistent with the chemical notion that Th-O bonds ought to be more covalent in nature than the very ionic Th-F bonds in Th:LiSrAlF$_6$ or Th:CaF$_2$ due to the energy and size of O$^{2-}$ and F$^{-}$ valence orbitals. Therefore, it points to a general and intuitive difference between fluoride \emph{versus} oxide hosts, namely the degree of hybridization between Th valence orbitals and anion orbitals. An alternative electron conversion mechanism would be that radiation (in particular $\beta ^-$), produced by the decaying nuclei in the sample, excites electrons from the valence band to the Th-related in-gap states, so that these electrons can then be excited to the conduction band upon decay of the isomer. However such a mechanism would in principle also occur for LiSrAlF$_6$, where such Th-related in-gap states are also predicted to occur even for the lowest-energy Th configurations \cite{elwell-2024}, and yet the radiative decay is clearly observed in LiSrAlF$_6$ in the present work.

\section{\label{sec:energy}Wavelength Determination}
\subsection{Calibration Procedure}

Calibration scans with a 250\,$\mu$m slit size covering a wavelength range of 120--180\,nm (see Ref.\,\cite{kraemer-2023-vuvsetup}) were taken throughout the experimental campaign.

Since the plasma source used for wavelength calibration is 36\,cm from the entrance slit and the implanted CaF$_2$ 350 crystal is 5.5\,mm from the slit, VUV photons from the plasma source and the radioactive source enter under different angles with respect to the optical axis. To eliminate a possible angular dependence of the wavelength calibration procedure, a MgF$_2$ diffuser crystal was mounted on the wheel to diffuse the light from the plasma  source at the position of  the implanted crystals. 
The three most prominent oxygen and nitrogen peaks were isolated and fitted separately to deduce their centroids in terms of motor position. While each peak appeared as a singlet at the resolution of the spectrometer in the $1^\mathrm{st}$ diffraction order, each peak was a multiplet containing two or more peaks which could be partially resolved in the second diffraction order. The fit procedure included each peak within the multiplet using experimentally deduced constraints on the relative peak positions and their intensities.

The wavelengths of the single lines used in the calibration are 130.2168\,nm, 141.194\,nm, 149.4675\,nm, and 174.2729\,nm, and a full list of spectral lines can be found in Ref.\,\cite{kraemer-2023-vuvsetup}. The linear fit from the wavelength calibration measurements performed at ISOLDE resulted in a slope of $b=115.24 \pm 0.08$
motor positions per nanometer. This corresponds to $0.008678\pm0.000006$\,nanometers per motor position which is within $1\sigma$ of the value reported in Ref.\,\cite{kraemer-2023-vuvsetup}.

\begin{figure}[h]
    \centering
    \includegraphics[width=0.48\textwidth]{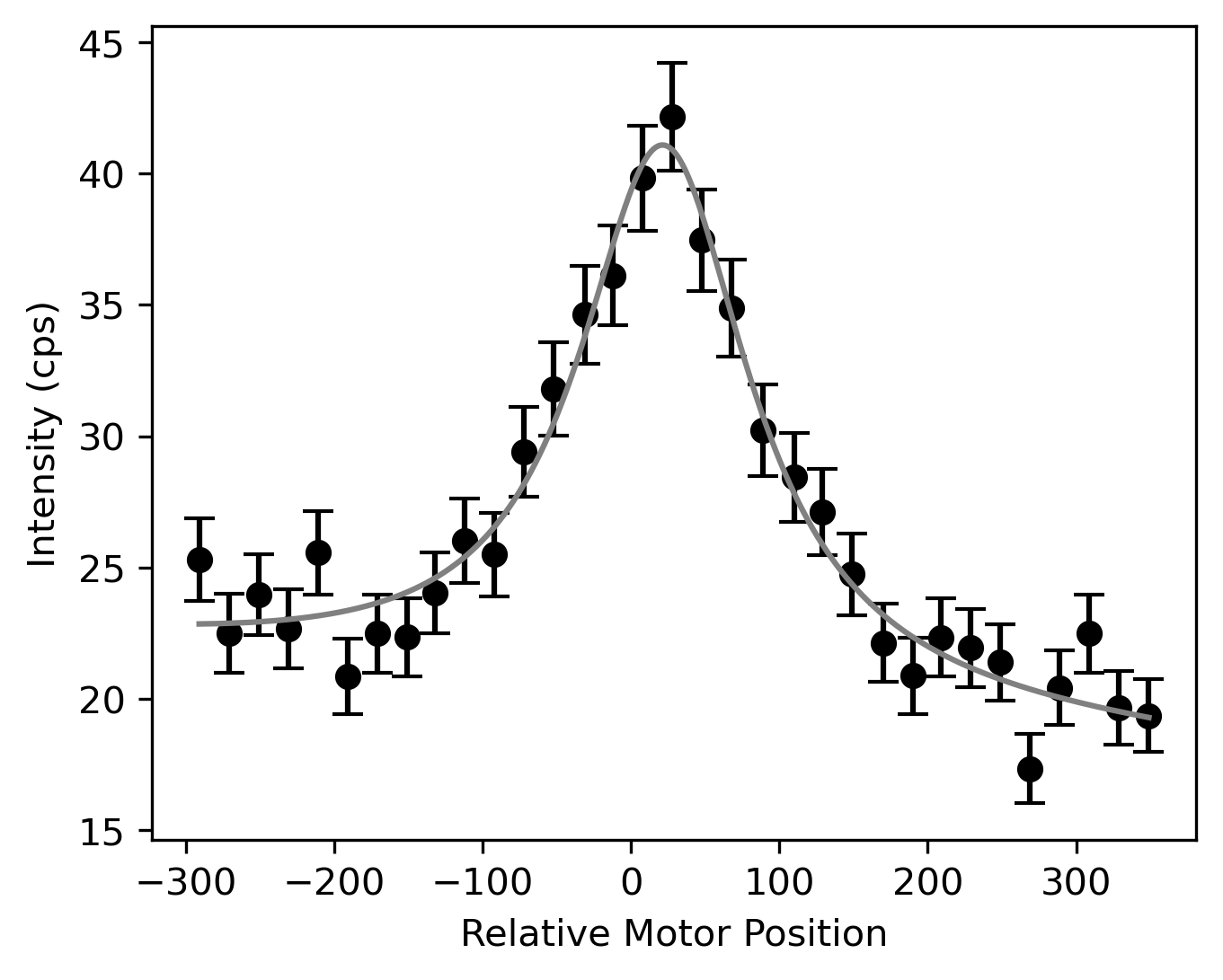}
    \caption{An example spectrum of the nitrogen line at 141\,nm used for the offset determination. The solid line is the fit to the data.}
    \label{fig:nitrogen-141}
\end{figure}

The single 141\,nm (141.194\,nm) nitrogen line was measured before and after each group of isomer wavelength measurements and used as an offset. Only scanning the 141\,nm nitrogen line allowed more calibration measurements to be performed in a shorter time period, resulting in better control over fluctuations in centroid position and compensating for any drifts of the system. A Lorentzian profile with a linear background, shown in Fig.\,\ref{fig:nitrogen-141}, was used to fit the calibration line due to the invading tails of the neighboring 131\,nm and 149\,nm multiplets. In the final wavelength determination of the isomer, the offset parameter is represented by $M_{141}$ in Eq.\,\ref{eq:isomerwave}.

\subsection{Uncertainty Analysis}
\paragraph{Offset determination:}
The nitrogen calibration line used as the offset consists of a triplet of three closely spaced transitions of neutral nitrogen \cite{Kaufman-1967,goldbach-1988,moore-1975}. The wavelength of this triplet from NIST is reported as 141.194\,nm, where the uncertainty is implied by a factor of $25\%$ on the least significant digit \cite{NIST,moore-1975}, which corresponds to 0.001\,nm. However, Ref.\,\cite{Kaufman-1967} shows that each of the nitrogen lines, given in Tab.\,\ref{tab:triplet-141}, were determined to a precision of 0.0001\,nm and a range of $\Delta\lambda=0.0018$\,nm between the lowest and highest wavelengths of the triplet. The range of the triplet as well as the uncertainty of the individual calibration lines were added in quadrature to give a conservative uncertainty of 0.002\,nm, which corresponds to 0.2\,motor positions.

\begin{table}[h]
\caption{\label{tab:triplet-141}
Nitrogen 141\,nm calibration line (Multiplet $uv10$) from Refs.\,\cite{Kaufman-1967,goldbach-1988}.}
\begin{ruledtabular}
\begin{tabular}{ccc}
Wavelength (nm) & Lower & Upper\\
\hline
\rule{-3pt}{3ex}
$141.19318\pm 0.00010$ & $2s^{2}2p^{3}$ $^{2}P_{1/2}\degree$ & 
$2s^{2}2p^{2}3s$ $^{2}D_{3/2}$\\
$141.19395\pm 0.00010$ & $2s^{2}2p^{3}$ $^{2}P_{3/2}\degree$ & 
$2s^{2}2p^{2}3s$ $^{2}D_{3/2}$\\
$141.19494\pm 0.00010$ & $2s^{2}2p^{3}$ $^{2}P_{3/2}\degree$ & 
$2s^{2}2p^{2}3s$ $^{2}D_{5/2}$\\
\end{tabular}
\end{ruledtabular}
\end{table}
\begin{table*}[ht]
\caption{\label{tab:wavelength-values-tabulated}
Isomer wavelength results for each series of wavelength determination measurements. $N_\mathrm{imp}$} is the implantation number for measurements taken at a 250\,$\mu$m slit and $t_\mathrm{imp}$ is the time elapsed implanting onto the crystal. $M_{141}$ is the centroid for the 141\,nm calibration line in motor positions. The uncertainty is the standard deviation between the measurements and the value reported is the average.
\begin{ruledtabular}
\begin{tabular}{cccccc}
$N_\mathrm{imp}$ & $t_\mathrm{imp}$ (s) & Crystal & $M_{141}$ (MP) & $M_{229m}$ (MP) &  $\lambda_{229m}$ (nm)\\
\hline
1 & 2233.50 & CaF$_2$ 350 & $-10834.6\pm 2.7$  & $-9995.1\pm2.6$ & $148.479\pm0.033$ \\
2 & 2132.15 & CaF$_2$ 350 & $-10835.7\pm 2.0$  & $-9993.4\pm4.3$ & $148.503\pm0.041$ \\
3 & 5486.23 & CaF$_2$ 350 & $-10831.7\pm 3.4$  & $-9988.6\pm2.5$ & $148.510\pm0.037$ \\
4 & 3591.94 & CaF$_2$ 350 & $-10828.6\pm 2.7$  & $-9989.4\pm3.7$ & $148.476\pm0.040$ \\
5 & 3669.53 & LiSrAlF$_6$ & $-10833.4\pm 1.9$  & $-100002.7\pm9.7$ & $148.402\pm0.086$ \\
\end{tabular}
\end{ruledtabular}
\end{table*}

\paragraph{Alignment with respect to the optical axis:}

The position of the radioactive source with respect to the optical axis influences the wavelength determination as VUV light can enter the spectrometer under an, on average, non-zero angle. To evaluate this effect, the position was estimated by comparing the experimental radiative decay fraction for the CaF$_2$ 350 crystal at a 250 $\mu$m slit size relative to the one at 2 mm, revealing a value that varied between 4.5\% and 8.1\% in the course of the experiment. This points to a potential deviation of the center of the source with respect to the optical axis of 2.5 mm. The extent of this effect on the wavelength determination was simulated using raytracing. This allowed to estimate the effect of a non-centered radioactive source as well as an off-center alignment of the entrance slit of less than 0.5 mm.
Revealing a maximum shift in wavelength of 0.15 nm which dominates the total systematic uncertainty.

\subsection{Wavelength Determination}
The isomer wavelength for each implantation was determined using Eq.\,\ref{eq:isomerwave}, where $\lambda_{141}$ is the nitrogen calibration line wavelength in nanometers, $b$ is the slope determined from the calibration measurements, $M_{141}$ is the 141\,nm nitrogen calibration line offset motor position, and $M_{229m}$ is the isomer motor position. The values used for wavelength determination are shown in Tab.\,\ref{tab:wavelength-values-tabulated}. A weighted average wavelength value of $148.491\pm0.019_\mathrm{(stat.)}\pm0.15_{\mathrm{(syst.)}}$ nm for CaF$_2$ and $148.402\pm0.086_{(\mathrm{stat.})} \pm0.15_{(\mathrm{syst.})}$\,nm for LiSrAlF$_6$ are reported.  
\begin{equation}
    \label{eq:isomerwave}
    \lambda_{229m} = \lambda_{141} - b\left(M_{141}-M_{229m}\right)
\end{equation}

\section{Conclusion}
In testing several crystal materials the signal from the radiative decay of $^{229m}$Th was observed in all but AlN and SiO$_2$. The AlN case follows the general understanding that a band gap smaller than the isomer transition energy facilitates the internal conversion process. However, such a mechanism does not necessarily apply to SiO$_2$. DFT calculations indicate that the absence of a radiative decay in SiO$_2$ may be due to internal conversion accelerated by the hybridization between Th valence orbitals and O$^{2-}$. 
Production rates were determined for each implantation and verified in an offline $\alpha$-decay analysis on the implanted crystals after the experiment, ensuring control over the amount of implanted isotopes and the purity of the beam.
Relative radiative decay fractions were obtained and the CaF$_2$ bulk crystal was determined to be the most efficient.

The measured emission wavelength of the radiative decay of the $^{229m}$Th isomer was determined to be $\lambda_{229m}=148.49\pm 0.02_\mathrm{(stat.)}\pm0.15_\mathrm{(syst.)}$ nm, improving a previous measurement using the same technique by a factor of 2.5 and in agreement with an independent recent determination \cite{kraemer-2023,hiraki-2024}. Using the spectroscopic results from photon emission of the isomer's radiative decay, laser excitation of $^{229m}$Th isomer became feasible and has meanwhile resulted in multiple observations \cite{tiedau-2024,elwell-2024,chang-2020}. This approach measures the wavelength of photon absorption when driving the ground- to isomeric nuclear transition. Laser excitation using broadband dye laser systems allowed to determine the absorption wavelength with a precision 0.0005 nm in CaF${}_2$, 0.0003 nm in LiSrAlF${}_{6}$ and frequency comb spectroscopy narrowed this value down to $\approx10^{-10}$ nm in CaF${}_2$. Within uncertainties, the results of emission spectroscopy presented in this work agree with the absorption wavelength obtained in laser excitation experiments.

The method of implanting an $A=229$ radioactive beam of $^{229m}$Th $\beta$-decay precursors into different large-bandgap materials and observing the isomer's radiative decay using a VUV spectrometer constitutes an efficient method to test various different materials for their potential use in a future solid-state nuclear clock without the need to grow doped crystals.

\begin{acknowledgments}
We acknowledge the support of the ISOLDE Collaboration and technical teams. Additionally, UW and JGC acknowledge support from the Portuguese Foundation for Science and Technology (Fundação para a Ciência e a Tecnologia FCT, CERN/FIS-TEC/0003/2021), and SB acknowledges support of the FWO, Belgium fellowship for fundamental research (contract n. 1167324N). This work is supported in part by Research Foundation Flanders (FWO, Belgium, nr. G078624N and I001323N), the FWO-FNRS under the EOS program project 40007501,the Special Research Fund (BOF KU Leuven C14/22/104), the European Union’s Horizon Europe Framework research and innovation programme under grant agreement no. 101057511 (EURO-LABS), the European Union’s Horizon 2020 research and innovation programme under the Marie Skłodowska-Curie grant agreement no. 101026762 and 861198 (LISA), the European Research Council (ERC ThoriumNuclearClock, grant agreement no. 856415), the European Union’s Horizon 2020 research and innovation programme under grant agreement no. 819957, the Romanian IFA grant CERN/ISOLDE, the Nucleu project no. PN 23 21 01 02, and the Bavaria California Technology Center (BaCaTeC 7 [2019-2]).

\end{acknowledgments}

\bibliography{apssamp}

\end{document}